\documentclass{emulateapj}

\usepackage{amssymb,amsmath,natbib,threeparttable,rotating}
\usepackage[graphicx]{realboxes}
\usepackage{ulem,color}

\gdef\omit#1{}

\def\sc{S\'ersic}
\def\req{$r_e=0.63\pm0.18$kpc}
\def\resf{$r_e=2.0\pm0.60$kpc}
\def\reqa{$r_e=0.57\pm0.18$kpc}
\def\reqs{$r_e=0.85\pm0.35$kpc}
\def\resfs{$r_e=2.6\pm1.2$kpc}
\def\nqs{$n_{sersic}=4.17\pm0.90$}
\def\nsfs{$n_{sersic}=2.18\pm2.03$}
\def\rzq{$r_e=5.08\pm0.28(1+z)^{-1.44\pm0.08}$kpc}
\def\rzsf{$r_e=6.02\pm0.28(1+z)^{-0.72\pm0.05}$kpc}
\def\fqnul{$6.0\pm1.7\times$}
\def\fqtwo{$1.9\pm0.7\times $}
\def\fsfnul{$2.4\pm0.7\times$}
\def\fsfq{$3.2\pm1.3\times$}
\def\mlim{$\mathrm{log_{10}}(M/M_{\sun})\geq 10.55$}
\def\smaxq{$\langle\Sigma\rangle_{max}=3.3\pm1.1\times10^{10}\mathrm{M_{\sun}kpc^{-2}}$}
\def\smaxsf{$\langle\Sigma\rangle_{max}=0.3\pm0.1\times10^{10}\mathrm{M_{\sun}kpc^{-2}}$}
\def\smaxqtwo{$\langle\Sigma\rangle_{max}=1.7\pm0.3\times10^{10}\mathrm{M_{\sun}kpc^{-2}}$}

\begin{document}

\title{The sizes of massive quiescent and star forming galaxies at $\mathrm{z\sim 4}$ with ZFOURGE\altaffilmark{11} and CANDELS}
\author{Caroline M. S. Straatman\altaffilmark{1,2}, Ivo Labb\'e\altaffilmark{2}, Lee R. Spitler\altaffilmark{3,4}, Karl Glazebrook\altaffilmark{5}, Adam Tomczak\altaffilmark{6}, Rebecca Allen\altaffilmark{5}, Gabriel B. Brammer\altaffilmark{7}, Michael Cowley\altaffilmark{3,4}, Pieter van Dokkum\altaffilmark{8}, Glenn G. Kacprzak\altaffilmark{5}, Lalit  Kawinwanichakij\altaffilmark{6}, Nicola Mehrtens\altaffilmark{6}, Themiya Nanayakkara\altaffilmark{5}, Casey Papovich\altaffilmark{6}, S. Eric Persson\altaffilmark{9}, Ryan F. Quadri\altaffilmark{6}, Glen Rees\altaffilmark{4}, Vithal Tilvi\altaffilmark{6}, Kim-Vy H. Tran\altaffilmark{6}, Katherine E. Whitaker\altaffilmark{10}}
\altaffiltext{1}{straatman@strw.leidenuniv.nl}
\altaffiltext{2}{Leiden Observatory, Leiden University, PO Box 9513, 2300 RA Leiden, The Netherlands}
\altaffiltext{3}{Australian Astronomical Observatory, PO Box 915, North Ryde, NSW 1670, Australia}	
\altaffiltext{4}{Department of Physics \& Astronomy, Macquarie University, Sydney, NSW 2109, Australia}
\altaffiltext{5}{Centre for Astrophysics and Supercomputing, Swinburne University, Hawthorn, VIC 3122, Australia}
\altaffiltext{6}{George P. and Cynthia W. Mitchell Institute for Fundamental Physics and Astronomy, Department of Physics and Astronomy, Texas A\& M University, College Station, TX 77843}
\altaffiltext{7}{Space Telescope Science Institute, 3700 San Martin Drive, Baltimore, MD 21218, USA}
\altaffiltext{8}{Department of Astronomy, Yale University, New Haven, CT 06520, USA}
\altaffiltext{9}{Carnegie Observatories, Pasadena, CA 91101, USA}
\altaffiltext{10}{Astrophysics Science Division, Goddard, Space Flight Center, Code 665, Greenbelt MD 20771, USA}

\altaffiltext{11}{This paper contains data gathered with the 6.5 meter Magellan Telescopes located at Las Campanas observatory, Chile.}

\begin{abstract}
We study the rest-frame ultra-violet sizes of massive ($\sim0.8\times10^{11}M_{\sun}$) galaxies at $3.4\leq z<4.2$, selected from the FourStar Galaxy Evolution Survey (ZFOURGE), by fitting single \sc\ profiles to HST/WFC3/F160W images from the Cosmic Assembly Near-Infrared Deep Extragalactic Legacy Survey (CANDELS). Massive quiescent galaxies are very compact, with a median circularized half-light radius \req. Removing $5/16$ (31\%) sources with signs of AGN activity does not change the result. Star-forming galaxies have \resf, \fsfq\ larger than quiescent galaxies. Quiescent galaxies at $z\sim4$ are on average \fqnul\ smaller than at $z\sim0$ and \fqtwo\ smaller than at $z\sim2$. Star-forming galaxies of the same stellar mass are \fsfnul\ smaller than at $z\sim0$. Overall, the size evolution at $0<z< 4$ is well described by a powerlaw, with \rzq\ for quiescent and \rzsf\ for star-forming galaxies. Compact star-forming galaxies are rare in our sample: we find only $1/14\Rightarrow7\%$ with $r_e/(M/10^{11}M_{\sun})^{0.75}<1.5$, whereas $13/16\Rightarrow81\%$ of the quiescent galaxies is compact. The number density of compact quiescent galaxies at $z\sim4$ is $1.8\pm0.8\times10^{-5}\mathrm{Mpc^{-3}}$ and increases rapidly, by $>5\times$, between $2<z<4$. The paucity of compact star-forming galaxies at $z\sim4$ and their large rest-frame ultra-violet median sizes suggest that the formation phase of compact cores is very short and/or highly dust obscured.

\end{abstract}

\section{Introduction}

In recent years massive quiescent galaxies have been found beyond $z=3$ \citep[e.g.][]{Chen04,Wiklind08,Mancini09,Fontana09,Marchesini10,Guo13,Stefanon13,Muzzin13,Spitler14} and even at $z\sim4$, when the universe was only 1.5 Gyrs old \citep{Straatman14}. Quiescent galaxies at high redshift ($z>1$) exhibit compact morphologies, with small effective radii \citep[e.g][]{Daddi05,vanDokkum08,Damjanov09}, which tend to become smaller with increasing redshift \citep{vanderWel14}. At $z\sim3$, they have sizes of $\sim1$kpc, $3-4\times$ smaller than early-type galaxies of similar stellar mass at $z\sim0$ \citep{Shen03,Mosleh13} and $2-3\times$ smaller than star-forming galaxies at the same redshift.

How compact quiescent galaxies are formed is still unclear. Simulations propose mechanisms in which gas-rich major mergers can induce central starbursts, resulting in a compact merger remnant \citep{Hopkins09,Wellons14}, or in which massive star-forming clumps move to the centers if galaxy disks are unstable \citep{Dekel09,Dekel14}. Alternatively they may have formed in a more protracted process at high redshift, when the universe was more dense \citep{Mo98}.

To understand these scenarios it is necessary to identify compact quiescent galaxies and their progenitors at the highest redshifts. Compact star-forming galaxies been found in small numbers at $z=2-3$ \citep{Barro14a,Barro14b,Nelson14}, but many host AGN, complicating the interpretation of the observations. At the same time, rest-frame ultra-violet (UV) or optically measured sizes of star-forming galaxies may be affected by dust-obscured central regions, thereby increasing their effective radii.

In this work we investigate the sizes of a stellar-mass complete sample of star-forming and quiescent galaxies at $z\sim4$. Throughout, we assume a standard $\mathrm{\Lambda CDM}$ cosmology with $\mathrm{\Omega_M=0.3,\ \Omega_{\Lambda}=0.7}$ and $H_0=70\mathrm{km\ s^{-1} Mpc^{-1}}$. The adopted photometric system is AB.

\section{Sample selection}\label{sec:data}

The galaxies were selected using deep $K_s$-band images from the FourStar Galaxy Evolution Survey (ZFOURGE; Labb\'e et al. in prep.), a near-IR survey with the FourStar Infrared Camera \citep{Persson13}, covering three $\mathrm{11'\times 11'}$ pointings, located in the fields CDFS \citep{Giacconi02}, COSMOS \citep{Scoville07} and UDS \citep{Lawrence07}. The ZFOURGE $K_s$-band selected catalogs are at least $80\%$ complete down to $K_s=24.53,\ 24.74$ and $25.07$ mag in each field, respectively \citep{Papovich14}. Photometric redshifts and stellar masses were derived using 5 near-IR medium-bandwidth filters on FourStar ($J_{1},J_{2},J_{3},H_s,H_l$), which provide fine sampling of the age-sensitive Balmer/$\mathrm{4000\AA}$ break at $1.5<z<4$, in combination with public data over a wavelength range $0.3-8\micron$ \citep{Straatman14}. Here we make additional use of HST/WFC3/F160W data from CANDELS \citep{Grogin11,Koekemoer11,Skelton14}, to examine galaxy sizes and Spitzer/MIPS $\mathrm{24\micron}$ data from GOODS-South (PI: Dickinson), COSMOS (PI: Scoville) and SPUDS (PI: Dunlop) to measure infrared flux.

The galaxies in this work have photometric redshifts $3.4\leq z<4.2$, stellar masses of \mlim\ and $K_s$-band signal-to-noise (SNR) of SNR$>7$. They are separated into quiescent and star-forming according to their rest-frame $U-V$ versus $V-J$ colours \citep{Labbe05,Williams09,Spitler14}, yielding 19 quiescent and 25 star-forming galaxies \citep{Straatman14}. Of these, 34 have HST/WFC3/F160W coverage. One quiescent galaxy has SNR$<3$ in F160W and is not included. Another star-forming galaxy with a highly uncertain redshift solution was also rejected from the sample, along with two star-forming galaxies that appear to consist of two sources each in the higher resolution HST images. In total we study 16 quiescent and 14 star-forming galaxies. We include a control sample at $2\leq z<3.4$ (326 sources) at similar mass and SNR.

\section{Galaxy sizes from HST/WFC3 imaging}

\subsection{\sc\ fits}\label{sec:gf}

\begin{table*}
\caption{Properties of 16 quiescent and 14 star forming galaxies}
\begin{center}
\Rotatebox{90}{\begin{threeparttable}
\begin{tiny}
\begin{tabular}{l r r l l l l l l l l l l l l r}
\hline
\hline
ID & R.A. & Decl & z & $\mathrm{K_{s,tot}}$ & $\mathrm{H_{tot}}$\tnote{a} & $\mathrm{H_{Galfit}}$\tnote{b} & $\mathrm{SNR}_{F160W}\tnote{a}$  & $\mathrm{M/10^{11}}$ & $r_{KRON}$\tnote{a} & $r_{1/2,maj}$ & $r_e$ & $b/a$ & $n_{sercic}$ & $A_v$ & $24\micron$\tnote{c}\\
& (deg) & (deg) & & (mag)& (mag) & (mag) & & $\mathrm{(M_{\odot})}$ & ($"$) & ($"$) & (kpc) & & & & ($\mu$Jy)\\
\hline
QUIESCENT \\
\hline
ZF-CDFS-209 &  53.1132774 & -27.8698730 &   3.56 &   22.6 &   24.1 & 
  24.3 $\pm$   0.0 &   64.6 & 0.76 &   0.23 &   0.06 $\pm$  0.01 & 
  0.27 $\pm$  0.07 &   0.37 $\pm$  0.08 &   4.00 & 0.3 & 
     -0.9 $\pm$      3.5 \\
ZF-CDFS-403 &  53.0784111 & -27.8598385 & 3.660\tnote{d} &   22.4 &   23.7 & 
  23.5 $\pm$   0.0 &  118.0 & 1.15 &   0.22 &   0.12 $\pm$  0.03 & 
  0.82 $\pm$  0.18 &   0.85 $\pm$  0.05 &   7.78 $\pm$  0.94 & 0.8 & 
     99.8 $\pm$    148.5$^{\dagger}$ \\
ZF-CDFS-4719 &  53.1969414 & -27.7604313 &   3.59 &   23.4 &   25.2 & 
  25.2 $\pm$   0.1 &   33.5 & 0.45 &   0.23 &   0.12 $\pm$  0.03 & 
  0.60 $\pm$  0.14 &   0.48 $\pm$  0.08 &   1.88 $\pm$  0.84 & 0.3 & 
      1.9 $\pm$      3.4 \\
ZF-CDFS-4907 &  53.1812820 & -27.7564163 &   3.46 &   23.6 &   25.0 & 
  25.1 $\pm$   0.1 &   38.2 & 0.40 &   0.28 &   0.08 $\pm$  0.02 & 
  0.56 $\pm$  0.13 &   0.86 $\pm$  0.12 &   3.28 $\pm$  0.90 & 0.8 & 
      1.4 $\pm$      3.6 \\
ZF-CDFS-5657 &  53.0106506 & -27.7416019 &   3.56 &   23.0 &   24.6 & 
  24.2 $\pm$   0.1 &   26.7 & 0.76 &   0.33 &   0.52 $\pm$  0.16 & 
  3.22 $\pm$  0.93 &   0.72 $\pm$  0.11 &   4.45 $\pm$  0.98 & 0.3 & 
      1.7 $\pm$      3.8$^{\dagger}$ \\
ZF-CDFS-617 &  53.1243553 & -27.8516121 & 3.700\tnote{d} &   22.3 &   23.5 & 
  23.5 $\pm$   0.0 &  135.1 & 0.69 &   0.22 &   0.10 $\pm$  0.02 & 
  0.55 $\pm$  0.11 &   0.59 $\pm$  0.03 &   4.00 & 0.3 & 
     86.3 $\pm$      3.4$^{\dagger}$* \\
ZF-COSMOS-13129 & 150.1125641 &   2.3765368 &   3.81 &   23.6 &   25.2 & 
  24.9 $\pm$   0.1 &   10.8 & 1.78 &   0.46 &   0.52 $\pm$  0.13 & 
  2.15 $\pm$  0.48 &   0.34 $\pm$  0.08 &   0.56 $\pm$  0.24 & 0.6 & 
    110.1 $\pm$     10.2* \\
ZF-COSMOS-13172 & 150.0615082 &   2.3786869 &   3.55 &   22.4 &   24.4 & 
  24.4 $\pm$   0.1 &   37.2 & 1.45 &   0.27 &   0.08 $\pm$  0.02 & 
  0.49 $\pm$  0.12 &   0.64 $\pm$  0.13 &   3.94 $\pm$  1.11 & 0.6 & 
      2.7 $\pm$      7.6 \\
ZF-COSMOS-13414 & 150.0667114 &   2.3823516 &   3.57 &   23.4 &   25.4 & 
  25.4 $\pm$   0.1 &   14.0 & 0.44 &   0.32 &   0.20 $\pm$  0.06 & 
  0.83 $\pm$  0.29 &   0.34 $\pm$  0.14 &   1.51 $\pm$  1.00 & 0.2 & 
      7.1 $\pm$      8.7 \\
ZF-UDS-10684 &  34.3650742 &  -5.1488328 &   3.95 &   24.1 &   25.9 & 
  25.2 $\pm$   0.2 &    8.5 & 0.85 &   0.32 &   0.50 $\pm$  0.17 & 
  2.42 $\pm$  0.77 &   0.47 $\pm$  0.18 &   4.63 $\pm$  1.68 & 1.0 & 
      8.8 $\pm$     12.8 \\
ZF-UDS-11483 &  34.3996315 &  -5.1363320 &   3.63 &   23.6 &   26.0 & 
  25.9 $\pm$   0.2 &    8.9 & 1.02 &   0.35 &   0.11 $\pm$  0.05 & 
  0.52 $\pm$  0.25 &   0.43 $\pm$  0.24 &   4.59 $\pm$  2.01 & 1.0 & 
      1.8 $\pm$     10.2 \\
ZF-UDS-2622 &  34.2894516 &  -5.2698011 &   3.77 &   23.0 &   24.6 & 
  24.5 $\pm$   0.1 &   29.9 & 0.87 &   0.30 &   0.13 $\pm$  0.03 & 
  0.76 $\pm$  0.19 &   0.66 $\pm$  0.10 &   4.00 & 0.9 & 
     12.2 $\pm$     10.6 \\
ZF-UDS-3112 &  34.2904282 &  -5.2620673 &   3.53 &   23.2 &   24.9 & 
  24.9 $\pm$   0.1 &   25.7 & 0.43 &   0.30 &   0.07 $\pm$  0.02 & 
  0.39 $\pm$  0.13 &   0.66 $\pm$  0.19 &   4.00 & 0.0 & 
    -10.9 $\pm$     10.6 \\
ZF-UDS-5418 &  34.2937546 &  -5.2269468 &   3.53 &   23.3 &   24.9 & 
  24.9 $\pm$   0.1 &   20.7 & 0.44 &   0.30 &   0.07 $\pm$  0.02 & 
  0.50 $\pm$  0.14 &   0.83 $\pm$  0.17 &   4.00 & 0.5 & 
     48.4 $\pm$     10.6 \\
ZF-UDS-6119 &  34.2805405 &  -5.2171388 &   4.05 &   23.8 &   25.5 & 
  25.4 $\pm$   0.2 &   10.6 & 0.55 &   0.32 &   0.26 $\pm$  0.15 & 
  1.26 $\pm$  0.75 &   0.49 $\pm$  0.20 &   4.00 & 1.0 & 
    -12.5 $\pm$      8.7 \\
ZF-UDS-9526 &  34.3381844 &  -5.1661916 &   3.97 &   24.2 &   25.9 & 
  25.8 $\pm$   0.3 &   11.5 & 0.89 &   0.21 &   0.10 $\pm$  0.05 & 
  0.39 $\pm$  0.35 &   0.34 $\pm$  0.24 &   2.03 $\pm$  2.28 & 1.8 & 
     38.7 $\pm$      8.7$^{\dagger}$* \\
\hline
STACK & -  & - &   3.66 & - & - & - & - & 0.81 & - & - &  0.85 $\pm$  0.35
 & - &   4.14 $\pm$  0.71 & - & - \\
\hline
 \\
STAR FORMING \\
\hline
ZF-CDFS-261 &  53.0826530 & -27.8664989 &   3.40 &   23.2 &   24.2 & 
  24.5 $\pm$   0.1 &   27.1 & 1.07 &   0.40 &   0.61 $\pm$  0.14 & 
  3.54 $\pm$  0.80 &   0.62 $\pm$  0.06 &   1.21 $\pm$  0.25 & 1.9 & 
     12.1 $\pm$      4.4$^{\dagger}$ \\
ZF-CDFS-400 &  53.1025696 & -27.8606110 &   4.10 &   24.3 &   25.1 & 
  25.1 $\pm$   0.2 &   23.9 & 0.52 &   0.33 &   0.24 $\pm$  0.13 & 
  1.45 $\pm$  0.78 &   0.78 $\pm$  0.11 &   3.40 $\pm$  1.40 & 0.9 & 
     31.3 $\pm$      3.6$^{\dagger}$* \\
ZF-CDFS-509 &  53.1167717 & -27.8559704 &   3.95 &   24.2 &   25.1 & 
  25.0 $\pm$   0.0 &   29.1 & 0.41 &   0.25 &   0.31 $\pm$  0.06 & 
  1.55 $\pm$  0.32 &   0.52 $\pm$  0.05 &   0.51 $\pm$  0.17 & 1.0 & 
     -4.5 $\pm$      4.1 \\
ZF-COSMOS-12141 & 150.0815277 &   2.3637166 &   4.00 &   24.0 &   24.7 & 
  24.1 $\pm$   0.2 &   18.8 & 0.45 &   0.34 &   0.81 $\pm$  0.27 & 
  3.58 $\pm$  1.09 &   0.40 $\pm$  0.10 &   4.92 $\pm$  1.35 & 1.1 & 
      0.9 $\pm$      8.0 \\
ZF-COSMOS-3784 & 150.1817627 &   2.2390490 &   3.58 &   22.9 &   23.9 & 
  23.8 $\pm$   0.1 &   26.6 & 0.36 &   0.38 &   0.53 $\pm$  0.13 & 
  3.40 $\pm$  0.78 &   0.77 $\pm$  0.10 &   1.88 $\pm$  0.33 & 0.5 & 
     -2.4 $\pm$     10.2 \\
ZF-UDS-11279 &  34.3843269 &  -5.1402941 &   3.72 &   25.0 &   26.6 & 
  26.4 $\pm$   0.3 &    4.5 & 0.46 &   0.32 &   0.15 $\pm$  0.10 & 
  0.96 $\pm$  0.54 &   0.81 $\pm$  0.23 &   1.00 & 2.2 & 
     29.3 $\pm$     12.5 \\
ZF-UDS-4432 &  34.3581772 &  -5.2409291 &   3.76 &   23.8 &   24.5 & 
  24.2 $\pm$   0.2 &   17.5 & 0.83 &   0.37 &   0.75 $\pm$  0.39 & 
  3.61 $\pm$  1.74 &   0.46 $\pm$  0.11 &   4.27 $\pm$  1.65 & 1.5 & 
    669.0 $\pm$     10.7* \\
ZF-UDS-4449 &  34.3409157 &  -5.2405076 &   3.84 &   23.1 &   24.4 & 
  24.9 $\pm$   0.1 &   17.2 & 0.41 &   0.35 &   0.44 $\pm$  0.10 & 
  1.90 $\pm$  0.41 &   0.38 $\pm$  0.07 &   0.23 $\pm$  0.14 & 1.0 & - \\
ZF-UDS-4462 &  34.3408661 &  -5.2402906 &   3.92 &   23.0 &   24.0 & 
  24.0 $\pm$   0.1 &   27.9 & 0.39 &   0.26 &   0.39 $\pm$  0.09 & 
  2.09 $\pm$  0.45 &   0.60 $\pm$  0.08 &   1.69 $\pm$  0.27 & 0.8 & 
     22.6 $\pm$      9.4 \\
ZF-UDS-5617 &  34.3407745 &  -5.2240300 &   4.17 &   24.5 &   26.0 & 
  24.5 $\pm$   0.3 &    6.3 & 0.42 &   0.37 &   2.33 $\pm$  0.72 & 
 10.74 $\pm$  3.30 &   0.45 $\pm$  0.18 &   4.92 $\pm$  1.51 & 1.3 & 
      9.5 $\pm$      9.7 \\
ZF-UDS-8379 &  34.4104004 &  -5.1821156 &   3.77 &   23.8 &   25.2 & 
  25.2 $\pm$   0.1 &   14.0 & 0.65 &   0.25 &   0.30 $\pm$  0.07 & 
  1.50 $\pm$  0.34 &   0.50 $\pm$  0.09 &   0.52 $\pm$  0.28 & 2.6 & 
    355.8 $\pm$     25.0* \\
ZF-UDS-8399 &  34.4105759 &  -5.1825032 &   3.44 &   24.4 &   25.3 & 
  25.0 $\pm$   0.1 &   11.9 & 0.43 &   0.23 &   0.69 $\pm$  0.16 & 
  2.28 $\pm$  0.49 &   0.20 $\pm$  0.05 &   0.14 $\pm$  0.17 & 2.5 & 
    106.6 $\pm$     25.1* \\
ZF-UDS-8580 &  34.3544159 &  -5.1797152 &   4.07 &   23.7 &   24.6 & 
  24.7 $\pm$   0.1 &   19.8 & 0.66 &   0.26 &   0.36 $\pm$  0.08 & 
  1.82 $\pm$  0.37 &   0.54 $\pm$  0.05 &   0.18 $\pm$  0.09 & 1.1 & 
      7.1 $\pm$      8.4 \\
ZF-UDS-9165 &  34.3225441 &  -5.1713767 &   4.06 &   23.4 &   24.2 & 
  24.6 $\pm$   0.1 &   33.8 & 0.68 &   0.31 &   0.11 $\pm$  0.03 & 
  0.66 $\pm$  0.14 &   0.72 $\pm$  0.09 &   1.00 & 0.3 & 
     43.3 $\pm$     10.1* \\
\hline
STACK & -  & - &   3.84 & - & - & - & - & 0.55 & - & -&   2.62 $\pm$  1.15
 & - &   2.17 $\pm$  2.41 & - & - \\
\hline
\end{tabular}
\end{tiny}
\begin{tablenotes}
\item[a]F160W, SNR and circularized KRON radius ($r_{KRON}$) crossmatched from 3D-HST \citep{Skelton14,vanderWel14}
\item[b]GALFIT and 3DHST magnitudes are consistent within $0.05\pm0.03$ mag on average, with dispersion 0.24.
\item[c]$\dagger$: X$-$ray detection \citep{Xue11}; $^*$: $L_{IR}>7\times10^{12}L_{\sun}$.  
\item[d]$zspec$ \citep{Szokoly04}
\end{tablenotes}
\end{threeparttable}}
\end{center}	
\label{tab:sztab}
\end{table*}
 
Sizes and structural parameters were measured by fitting \sc\ \citep{Sersic68} profiles on $6\arcsec\times6\arcsec$ HST/WFC3/F160W image stamps using GALFIT \citep{Peng10}. In particular, we measure the half-light radius, encapsulating half the sources' integrated light. The corresponding parameter in GALFIT is the half-light radius along the semi-major axis ($r_{1/2,maj}$), which can be converted to circularized effective radius ($r_e=r_{1/2,maj}\sqrt(b/a)$), with $b/a$ the axis ratio.

We manually subtracted the background in each image stamp, masking sources and using the mode of the pixel flux distribution. Sky estimation in GALFIT was turned off. Neighbouring objects at $r>1.1\arcsec$ from the source were effectively masked by setting their corresponding pixels in the image to zero flux and increasing those in the noise image by $\times10^6$. Close neighbouring objects were fitted simultaneously.

We created mean PSFs for each field by stacking image stamps of bright stars (masking all neighbouring sources). As many of the galaxies are small we investigate the impact of PSF choice. We repeated the fitting using the hybrid PSF models of \citet{vanderWel12} and find marginally larger ($<5\%$) sizes. In particular, for the smallest galaxies ($r_e<0.20\arcsec$), we find a median $r_e/r_{e,PSFvdW}=0.93\pm0.05$. 

Errors on the individual measurements were calculated using a Monte Carlo procedure. After subtracting the best-fit GALFIT models from the sources, we shifted the residuals by a random number of pixels, added back the model and used this as input for GALFIT. Repeating this $>200\times$ for each galaxy, errors were calculated as the $1\sigma$ variation on these measurements. We report our results in Table \ref{tab:sztab}.

In the fits, the \sc\ index ($n_{sersic}$) was restricted to $0.1<n_{sersic}<8.0$. If $n_{sersic}$ reached the extreme value $0.1$ or $8.0$, GALFIT was rerun while forcing $n_{sersic}=1$ for star-forming and $n_{sersic}=4$ for quiescent galaxies. These values correspond to the median $n_{sersic}$ of galaxies with well-constrained fits and $\mathrm{SNR}_{F160W}>15$. 

At $z\sim4$, this happens for $6/16\Rightarrow38\%$ quiescent and $2/14\Rightarrow14\%$ star-forming galaxies. 
To explore systematic effects introduced by the choice of profile, we set $n_{sersic}=1.0$ or $n_{sersic}=4.0$ for bright ($mag_{F160W}(AB)<24.5$) and compact sources ($r_e<0.20\arcsec$) and find on average $r_{e,n=1}/r_{e,n=4}=0.80\pm0.13$, corresponding to a systematic uncertainty of $20\%$. We add this in quadrature to the uncertainties from the Monte Carlo procedure for each galaxy. Systematic biases of this level do not affect the main results. For comparison, \cite{vanderWel12} derived typical systematic uncertainties on size of $\sim12\%$ for faint F160W$=24-26$ and small $r_e<0.3\arcsec$ galaxies.

As many galaxies have modest SNR, we tested the reliability of our measurements by a simulation, in which we inserted source models, convolved with the instrument PSF, in the F160W images. These have adopted magnitudes of $25<mag_{F160W}(AB)<26$ and size of $0.06<r_e(\arcsec)<0.3$. 
We find $r_{e,out}/r_{e,in}=0.97\pm0.05$, with $r_{e,in}$ and $r_{e,out}$ the input and output effective radii, showing that we can recover the sizes of faint compact sources without bias. As an additional test we determine the size distribution of point sources, by inserting PSFs in the images and measuring their size. We can constrain the size of bright objects to $0.01\arcsec$ at 95\% confidence, which we adopt as a minimum uncertainty on the sizes.

We crossmatched our sample at $2\leq z<4.2$ with the size catalogs of \cite{vanderWel14}, based on the 3D$-$HST photometric catalogs \citep{Skelton14}. We find that the sizes and \sc\ indices agree well, with a median $r_{e,ZFOURGE}/r_{e,3DHST}=1.004\pm0.01$ and $n_{ZFOURGE}-n_{3DHST}=-0.012\pm0.058$. 

We test for color gradients between rest-frame UV sizes and rest-frame optical sizes, using a rest-frame color and stellar-mass matched control sample at $z\sim3$. We find F160W (rest-frame $4000\mathrm{\AA}$) sizes are $0\pm6\%$ and $6\pm11\%$ smaller than F125W (rest-frame $3000\mathrm{\AA}$) sizes for star-forming and quiescent galaxies, respectively.

\subsection{Stacking}

\begin{figure*}
\includegraphics[width=0.146\textwidth]{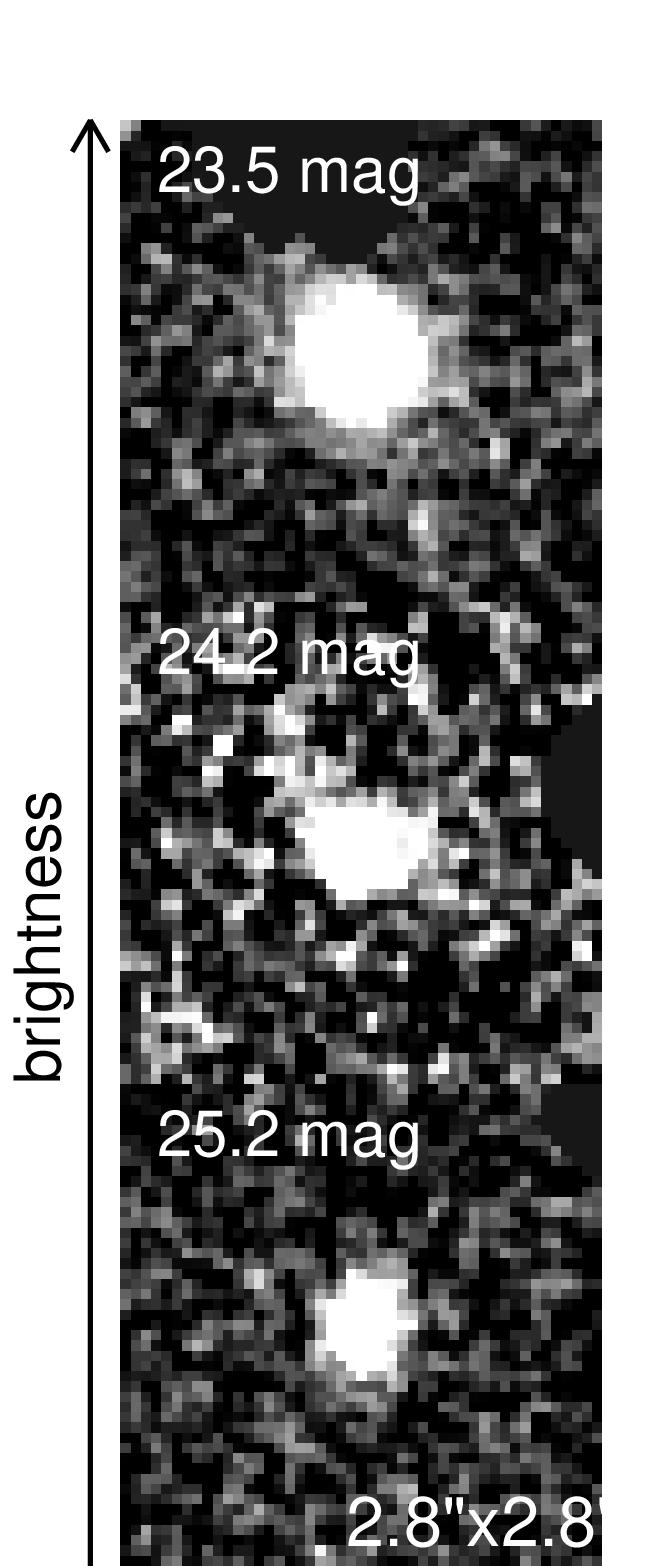}
\includegraphics[width=0.239\textwidth]{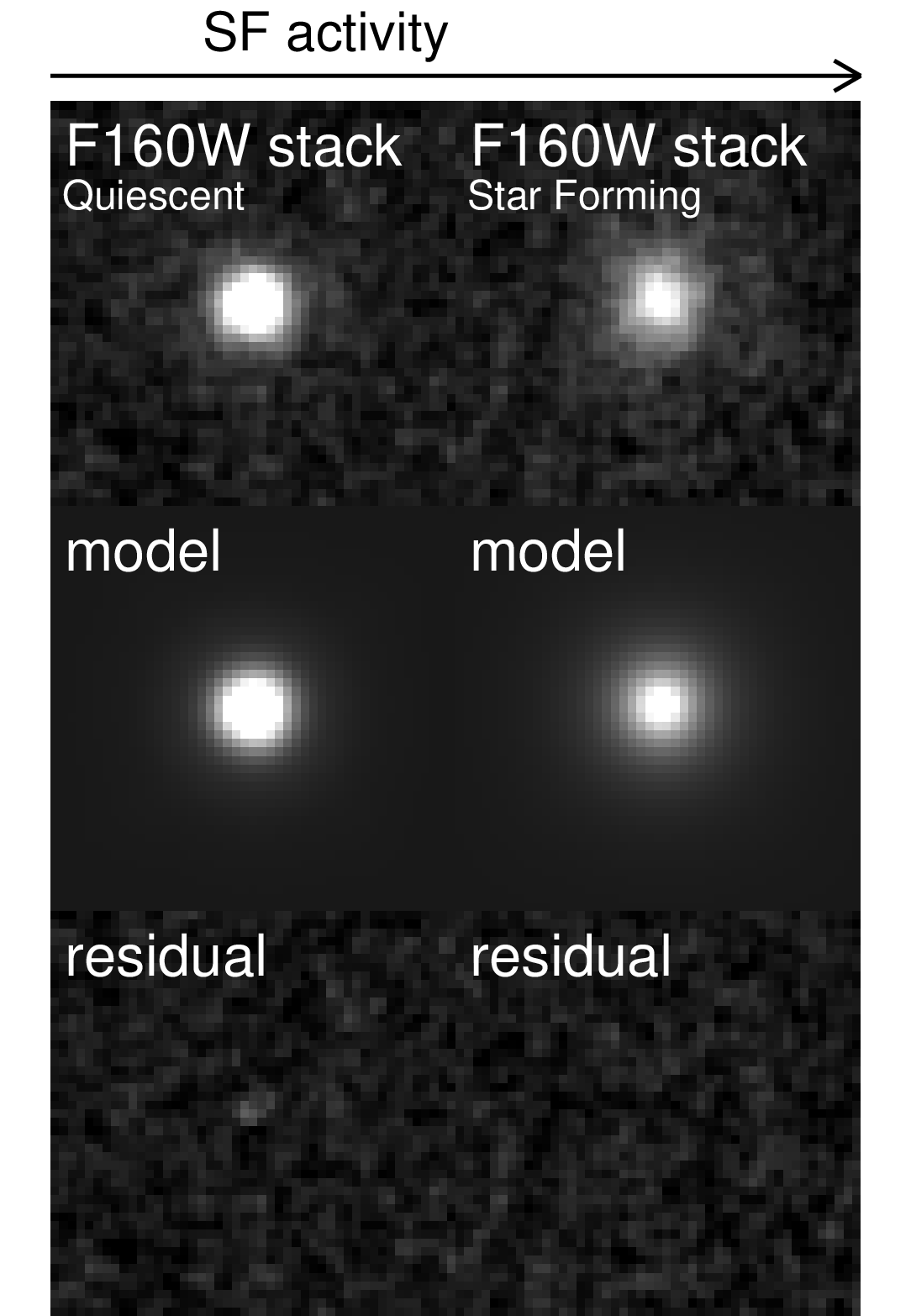}
\includegraphics[width=0.199\textwidth]{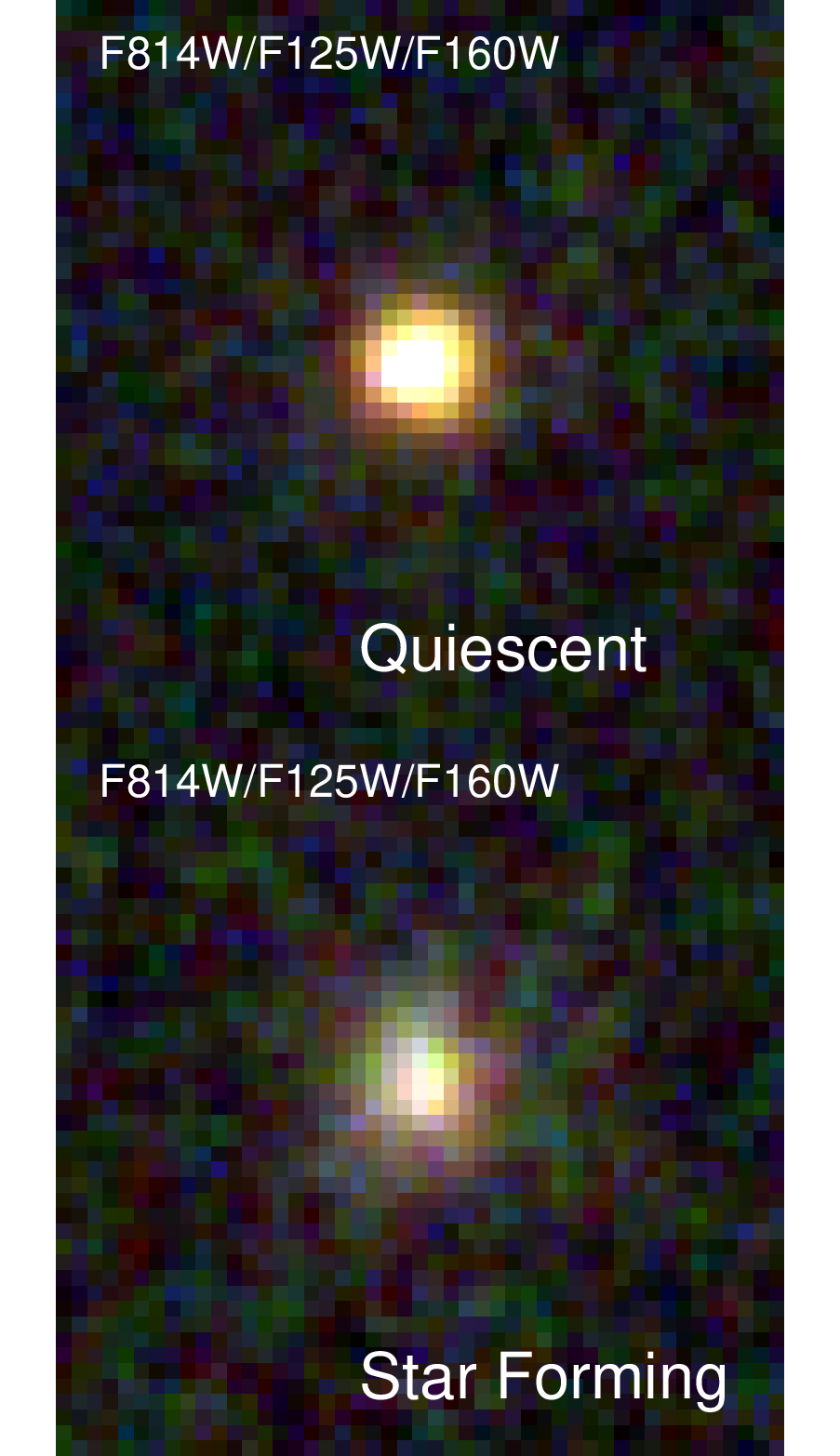}
\includegraphics[width=0.414\textwidth]{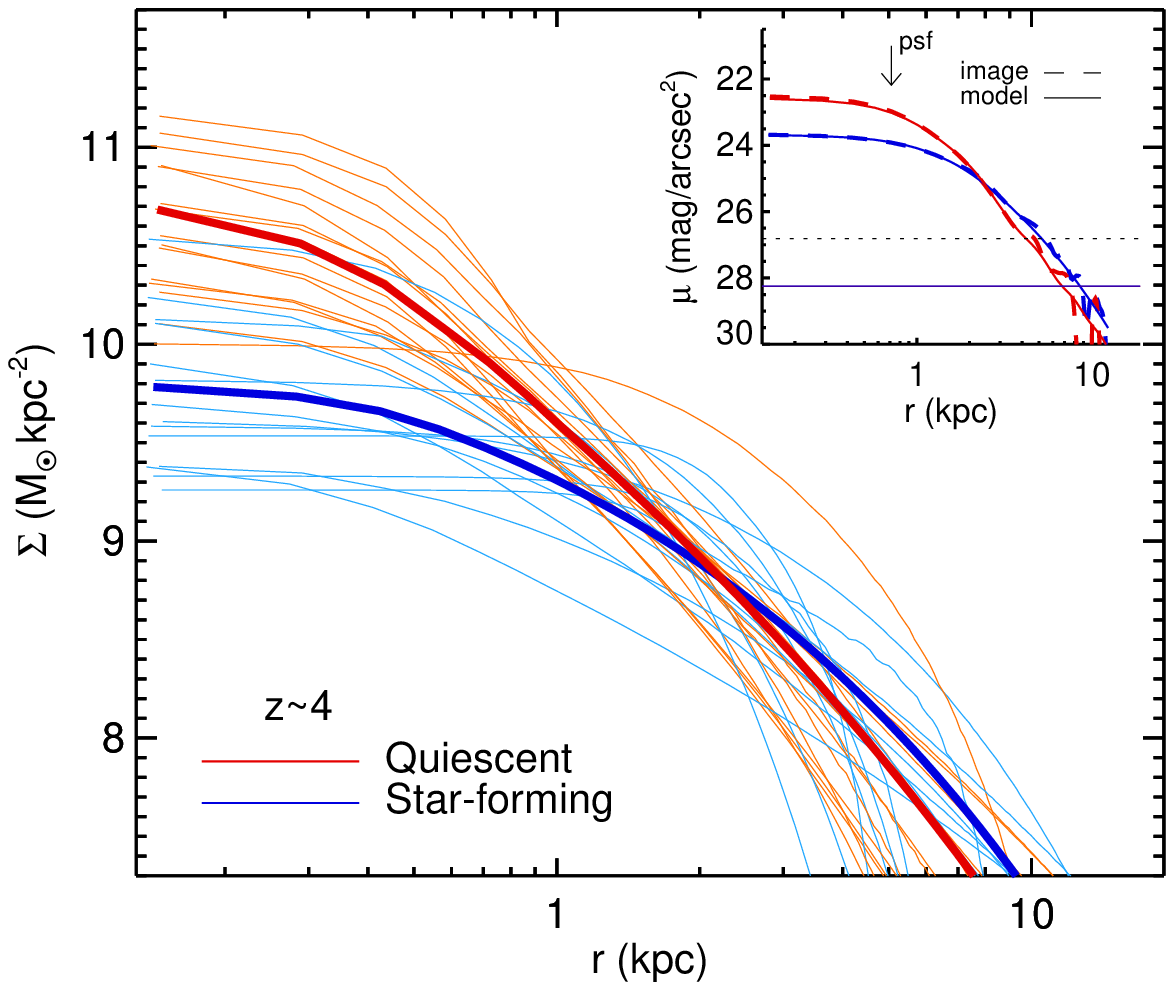}
\caption{Left: Example galaxies at $z\sim 4$ of varying magnitude. Second: stacks of the quiescent and star forming subsamples, with the corresponding best-fit models and residuals after subtracting the models. Third: F814W/F125W/F160W stack color composites. Right: Stellar mass surface density profiles. Thin orange and blue lines represent individual measurements of quiescent and star-forming galaxies, respectively. Thick lines represent the stacks. The inset shows the surface brightness profiles of the stacks, with horizontal lines indicating $3\sigma$ brightness limits of $28.3\ \mathrm{mag/arcsec^{2}}$, measured in annuli of $0.06\arcsec$ (0.43kpc) width at $r>1\arcsec$. The background limit for individual galaxies (dotted line) is $26.8\ \mathrm{mag/arcsec^{2}}$.}
\label{fig:fig1}
\end{figure*}

We also measure the average sizes by stacking the background subtracted image stamps of the two subsamples, normalizing each by mean stellar mass. Neighbouring sources were masked. The final stacks were obtained by calculating the mean value at each pixel location of the image stamps. 

We ran GALFIT using the same input parameters as for the individual galaxies. Errors were estimated by bootstrapping, i.e. randomly selecting galaxies, recreating the image stacks and rerunning GALFIT. 

In Figure \ref{fig:fig1} we show the stacks and examples of individual galaxies. The stack of quiescent galaxies is redder than the stack of star-forming galaxies and has a more compact morphology. We also show stellar mass surface density profiles ($\Sigma(\mathrm{M_{\Sun}/kpc^2})=M(<r)/(\pi r^2)$), obtained from the light profile measured in concentric apertures of radius $r$ and assuming a constant mass-to-light ratio. For the stacked profiles we used the mean mass of the galaxies in each stack. They are consistent with the individual profiles within the uncertainties, suggesting that the stack does not reveal an extended low surface brightness component, down to a surface brightness limit of $28.3\ \mathrm{mag/arcsec^{2}}$.

\subsection{Contamination by AGN}

A substantial fraction of sources show signs of AGN activity either from X$-$ray detections or strong $24\micron$ (rest-frame $5\micron$, tracing hot dust). As WFC3/F160W ($\lambda=1.5396\micron$) corresponds to rest-frame $2960-3500\mathrm{\AA}$ (UV) at $3.4\leq z<4.2$, it could be that an AGN is dominating their central light, leading to small sizes of the single \sc\ fits.

In the quiescent sample we find four X$-$ray detected galaxies, two of which are spectroscopically confirmed type-II QSOs \citep{Szokoly04}. Another has strong $24\micron$, which could either point towards dust-obscured star-formation or AGN activity. Several have small positive residuals after subtracting the best fit, suggesting the presence of a central point source. These $5/16$ (31\%) galaxies were re-fit with two components, a \sc\ model and a pointsource-like model (represented by a Gaussian with FWHM$=0.1$pixels) to trace possible AGN light. In these models, the point source accounts for $4.3-68\%$ of the total light (with $57$\% and $68$\% for the type-II QSOs, but on average $6.2$\% for the remaining 3 AGN candidates). The average size of the \sc\ component increases by $1.5\times$ (from a median $r_e=0.13\pm0.12\arcsec$ to $r_e=0.20\pm0.03\arcsec$). 

Amongst the star-forming galaxies two are X$-$ray detected, and four are very bright at $24\micron$ (L$>7\times10^{12}L_{\Sun}$ or $SFR>1200M_{\Sun}/yr$). Re-fitting with a two-component model attributes $0.9-39.4\%$ of the light to a point source, while the extended component changes in size by $0.65\times$ (from $r_e=0.31\pm0.15\arcsec$ to $r_e=0.19\pm0.02\arcsec$). We note that for the most extended sources, adding central light reduces the \sc\ index $n_{Sersic}$ of the extended component, and can result in a smaller $r_e$.

We additionally estimated the possible AGN contribution from the galaxy SEDs. We first determine the best fitting powerlaw bluewards of rest-frame $0.35\micron$ and at observed $8\micron$ \citep{Kriek09}. Then we fit the sum of the powerlaw and the original best-fit EAZY template to the data. The contribution of the AGN powerlaw template to F160W is $1.1-7.4\%$ for the 5 quiescent galaxies and $0.9-2.9\%$ for the 6 star-forming galaxies. 

While the two-component fits and SEDs indicate that a point source contribution is probably small, the true contribution and its effect on the sizes remain unclear. 

\section{Results}\label{sec:size-results}

\begin{figure*}
\includegraphics[width=\textwidth]{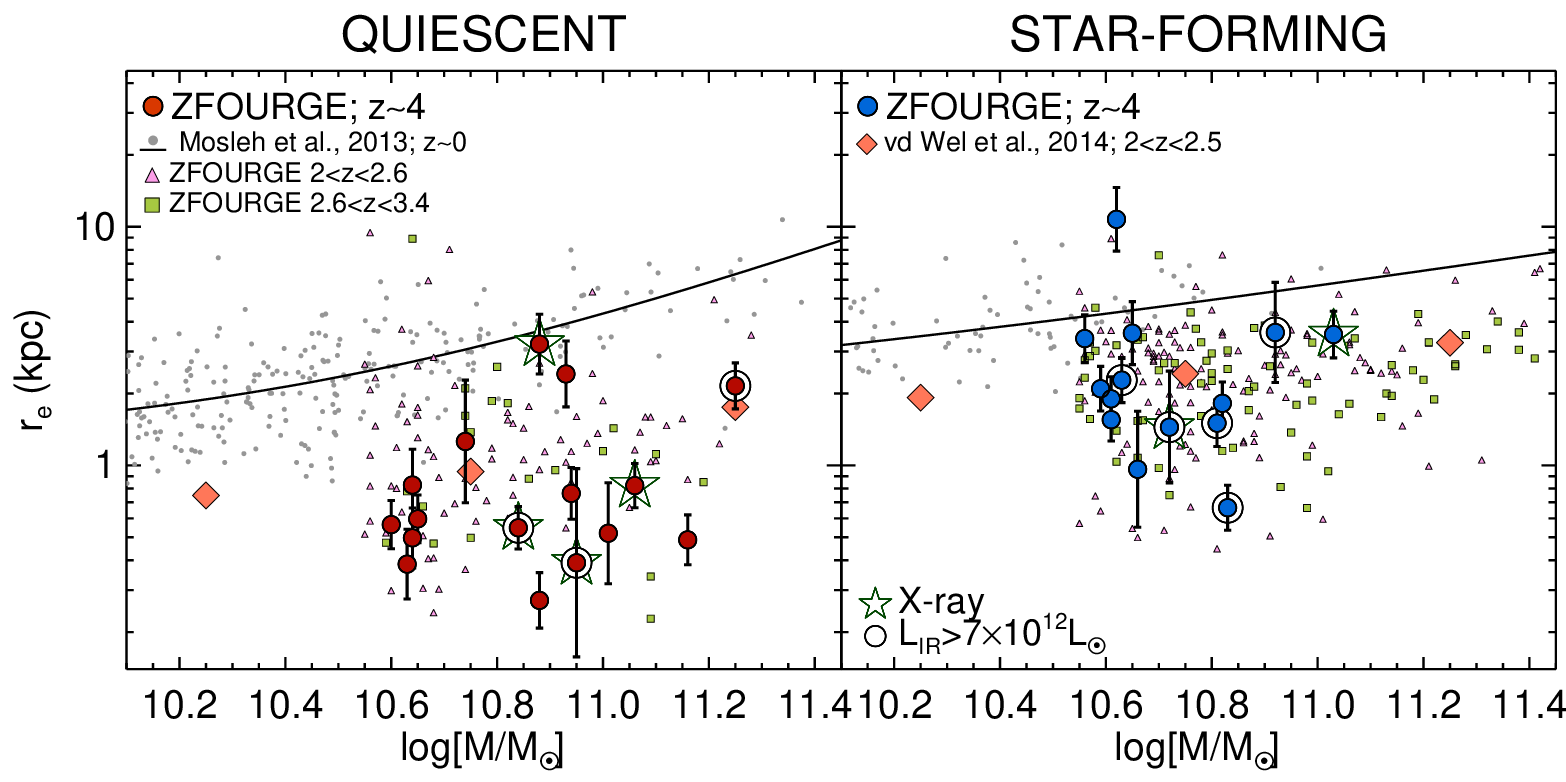}
\caption{Circularized effective radius for galaxies at $z\sim4$. In purple and green we show our control sample at $2\leq z<2.6$ and $2.6\leq z<3$ and in orange the median of \cite{vanderWel14} at $2<z<2.5$. The black solid line is the $z\sim0$ relation of \cite{Mosleh13}. X$-$ray detections and bright $24\micron$ sources are indicated with stars and open circles. The median sizes are \req\ (quiescent galaxies) and \resf\ (star-forming galaxies).}
\label{fig:msize}
\end{figure*}

We show the effective radius as a function of stellar mass in Figure \ref{fig:msize}. Quiescent galaxies at $z\sim4$ are very compact, with a bootstrapped median size \req. When we remove AGN we find a similar result: \reqa. 

Star-forming galaxies have \resf. They are \fsfq\ larger than quiescent galaxies. Both samples have a large spread in size, with some almost as large as at $z\sim0$, showing that at $z\sim4$ the population is already very diverse. On average the sizes lie well below the $z\sim0$ relation \citep{Mosleh13}, by \fqnul\ for quiescent and \fsfnul for star-forming galaxies. Quiescent galaxies are also \fqtwo smaller than at $2\leq z<2.2$.

\begin{figure*}
\includegraphics[width=\textwidth]{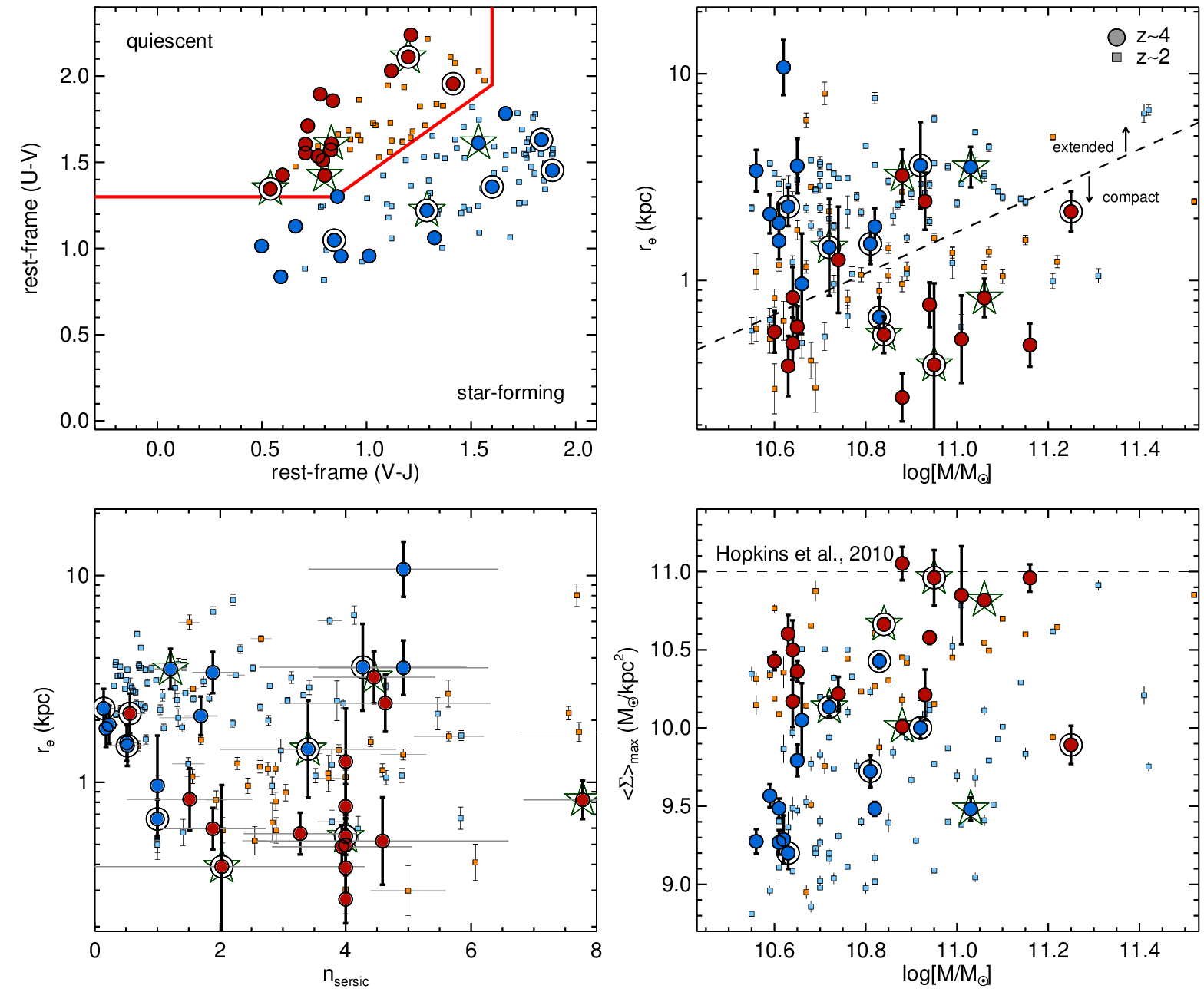}
\caption{Top-left: UVJ diagram of $z\sim4$ galaxies (symbols as in Figure \ref{fig:msize}). Small squares represent galaxies at $2.0\leq z<2.2$. Top-right: Stellar mass versus size. Bottom-left: \sc\ index versus size. Bottom-right: stellar mass versus maximum stellar mass density. The horizontal dashed line is the empirical limit of \cite{Hopkins10}. Only one $z\sim4$ star-forming galaxy is compact. On average quiescent galaxies have smaller sizes, higher \sc\ indices and higher central densities than star-forming galaxies.}
\label{fig:sb}
\end{figure*}
\vspace{50pt}

In Figure \ref{fig:sb} we show \sc\ index versus size for the $z\sim4$ galaxies, and a sample at similar mass at $2\leq z<2.2$. Star-forming galaxies have smaller \sc\ index, with on average $n_{sersic}=1.3\pm0.7$, compared to $n_{sersic}=3.2\pm1.2$ for quiescent galaxies.  The difference between the two populations is also clear from the stellar mass density profiles in Figure \ref{fig:fig1}, with quiescent galaxies having steeper profiles and more centralized flux. In Figure \ref{fig:sb} we also plot $\langle\Sigma\rangle_{max}$, defined as the average stellar mass density inside the radius where $\Sigma(\mathrm{M_{\Sun}/kpc^2})$ falls of by a factor of two \citep{Hopkins10}, with uncertainties from the Monte Carlo procedure described in section \ref{sec:gf}.

Quiescent galaxies at $z\sim4$ have a median \smaxq, much higher ($\sim10\times$) than for star-forming galaxies: \smaxsf, and more similar to $2\leq z<2.2$ quiescent galaxies: \smaxqtwo.

When stacking we find \reqs\ (quiescent) and \resfs\ (star-forming), and \sc\ indices \nqs\ and \nsfs, respectively. The effective radius of the quiescent stack is slightly larger than the median of the individual galaxies, by $1.3\pm0.3\times$ at $<1\sigma$ significance, but overall the results are consistent.

\begin{figure*}
\begin{center}
\includegraphics[width=0.9\textwidth]{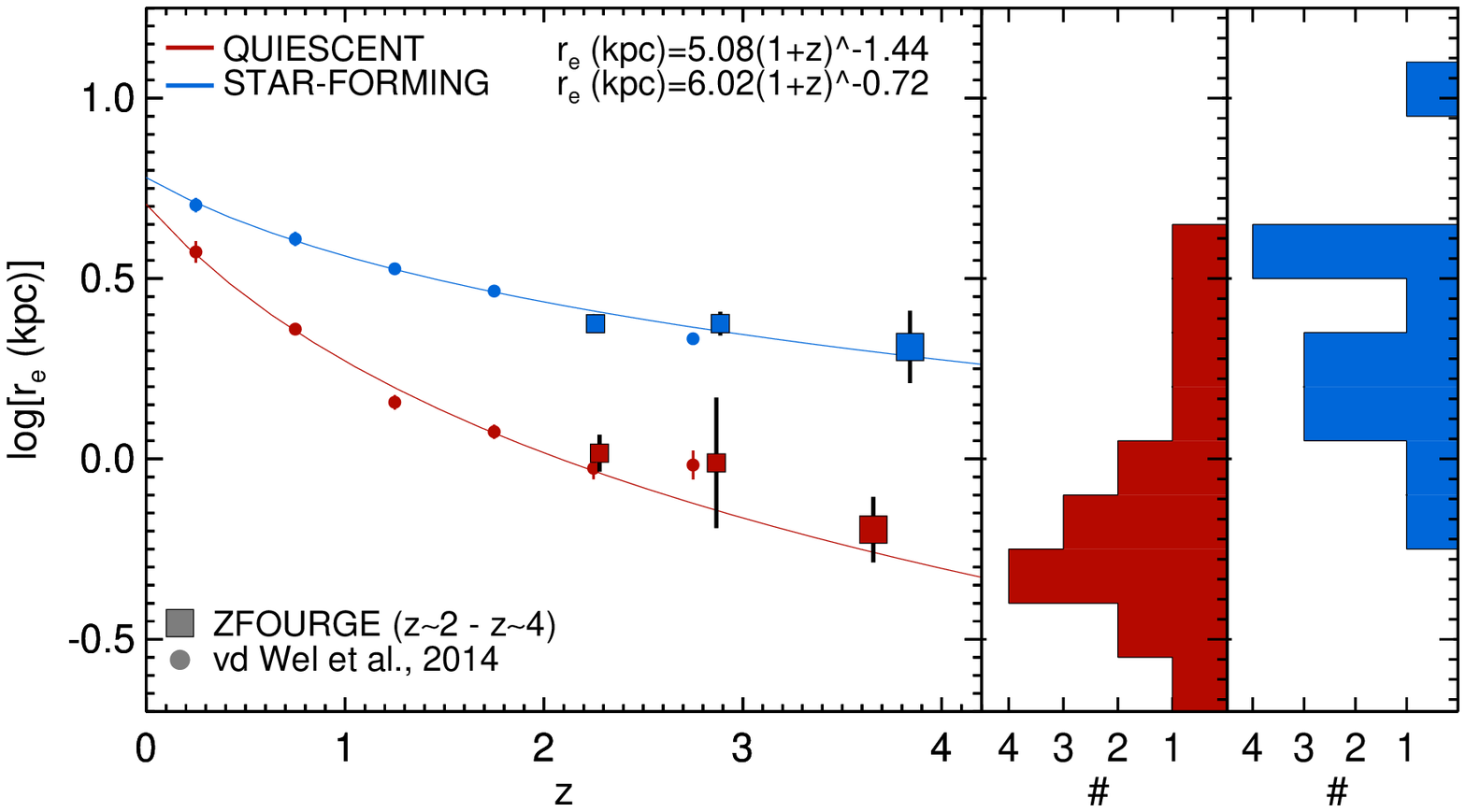}
\includegraphics[width=0.89\textwidth]{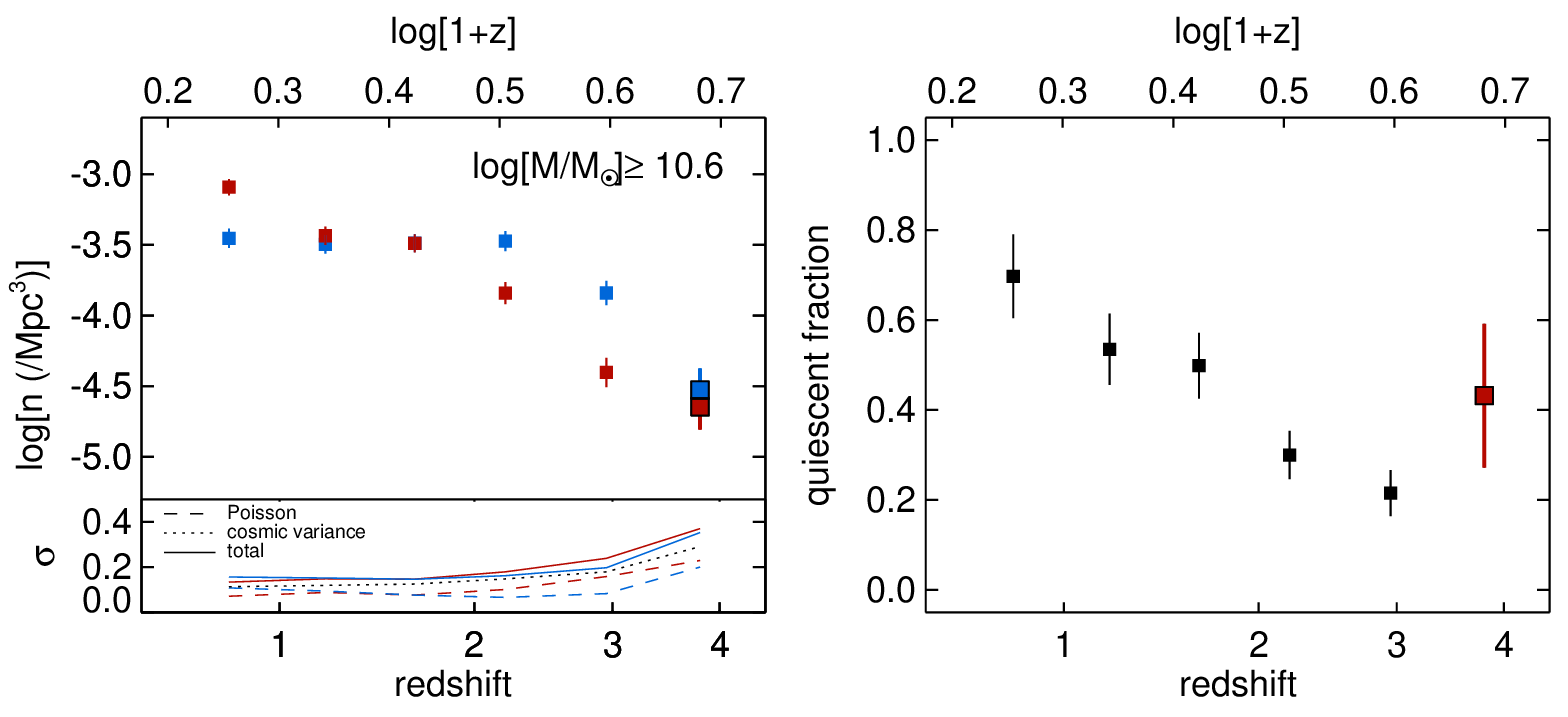}
\caption{Top: Effective radius versus redshift for galaxies with $10.5<\mathrm{log_{10}}(M/M_{\sun})<11.0$ at $2\leq z<3.4$ \citep{vanderWel14} and \mlim\ at $3.4\leq z<4.2$ (filled squares). Quiescent galaxies follow \rzq\ and star-forming galaxies \rzsf\ (solid curves). The histograms show the size distribution at $z\sim4$. Bottom: Number density (left) and quiescent fraction (right), including galaxies without HST coverage. In the left panel we include the relative Poissonian uncertainties and the effect of cosmic variance. The total uncertainty on number density increases to 40\% at $z\sim4$.}
\end{center}
\label{fig:rz}
\end{figure*}

In Figure \ref{fig:rz} we show the median sizes at the respective mean redshifts of the two subsamples. Comparing with lower redshift, they continue to follow a trend of decreasing size with increasing redshift. Our control sample of galaxies at $2\leq z<3.4$ with $10.5\leq\mathrm{log_{10}}(M/M_{\sun})<11$ corresponds well with the results of \cite{vanderWel14}, which suggest the same trend. 

We fit a relation of the form $r_e=A(1+z)^{B}$kpc at $0<z<4$, using the measurements of \cite{vanderWel14} at $z<2$. We find \rzq\ for quiescent and \rzsf\ for star-forming galaxies. We note that our sample at $z\sim4$ includes higher mass ($\mathrm{log_{10}}(M/M_{\sun})\geq 11$) galaxies. If we remove the most massive galaxies, we find the same evolutionary relation. 

To test for incompleteness for diffuse galaxies, we redshift a stellar-mass matched sample with $r>2$kpc and $n_{sersic}<2.5$ at $z\sim2.5$ to $z=3.7$ and find 70\% completeness.

\section{Discussion}	

Our results show that the galaxies at $z\sim4$ in this study obey similar relations between size and star-forming activity as galaxies at lower redshift.  Quiescent galaxies are compact, while star-forming galaxies are more extended and diffuse. The difference is also clear when selecting purely on size: if we define compactness as $r_e/(M/10^{11}M_{\sun})^{0.75}<1.5$ \citep{vanderWel14}, $13/14$ (93\%) of massive compact galaxies would be classified as quiescent, and $13/16$ (81\%) of larger galaxies as star-forming (Figure \ref{fig:sb}). 

The number density of compact, \mlim, quiescent galaxies at $z\sim4$ is $1.8\pm0.8\times10^{-5}\mathrm{Mpc^{-3}}$, increasing by $>5\times$ between $3.4\leq z<4.2$ and $2\leq z<2.2$, towards $1.0\pm0.3\times10^{-4}\mathrm{Mpc^{-3}}$. This suggests we are probing a key era of their formation, and we would expect to see their star-forming progenitors in abundance.

Small effective radii for star-forming galaxies have been reported at $z=2-3$ \citep{Barro14a,Barro14b,Nelson14}. They are rare in our sample: we find 1/14 with $r_e/(M/10^{11}M_{\sun})^{0.75}<1.5$. On average, star-forming galaxies at $z\sim4$ are twice as large as quiescent galaxies at $z\sim2$. If they are the direct progenitors of $z<4$ compact quiescent galaxies, we expect them to be similar, not only in size, but also in \sc\ index and central surface density \citep{Nelson14}. However, we find smaller $n_{Sersic}$ for star-forming galaxies, while the central densities indicate that they must increase in $\langle\Sigma\rangle_{max}$ by $5-10\times$, to match the more cuspy profiles of $z=2-4$ quiescent galaxies. 

In a recent simulation, \citet[Ilustris;][]{Wellons14} trace the evolution of galaxies to $z=2$. They indeed identified two theoretical formation tracks: one in which a brief and intense central starburst prompted by a gas-rich major merger causes the galaxies' half-mass radius to decrease dramatically. The second is that of a more gradual but early formation, with small galaxy sizes due to the higher density of the universe. In the second case, nearly all of the stellar mass is in place at $z>4$. 

Comparing with the observations, we find that 19/44 of massive $z\sim4$ galaxies are classified as quiescent, whereas all similarly massive galaxies in Illustris are still actively star-forming, with a typical $SFR=100-200M_{\Sun}/yr$. This level of star-formation is ruled out at $>3\sigma$ by Herschel observations of the $z\sim4$ quiescent galaxies \citep{Straatman14}. At the same time, the fraction of compact galaxies in our sample is 47\% ,versus $\sim20\%$ in Illustris. Hence massive galaxies appear to quench their star-formation earlier and to be more compact than in simulations. 

The paucity of compact star-forming galaxies at $z\sim4$  and their large median rest-frame UV size is puzzling. At face value it suggests that the rapid increase in number density of compact quiescent galaxies cannot be explained by simple shutdown of star-formation in typical star-forming galaxies of similar stellar mass. A possible solution is a rapidly forming dense core, i.e. a central starburst. Then the chance to observe the progenitors in our sample is small, as it is proportional to the duration of the main star-forming episode. For example, if compact cores of $2\leq z<2.2$ quiescent galaxies formed at random times between $2.5<z<6$, with a typical 100Myr central starburst duration, their predicted number density at $z\sim4$ would be $\sim6\times10^{-6}\mathrm{Mpc^{-3}}$. The observed number density of compact star-forming galaxies is $1.4\pm1.4\times10^{-6}\mathrm{Mpc^{-3}}$: smaller, but in a similar range given the large uncertainties.

We note that the remarkably high fraction of quiescent galaxies at $z\sim4$ (Figure \ref{fig:rz}) is still uncertain. Current limits on the average dust-obscured SFR are weak \citep[$<75M_{\Sun}/yr(3\sigma)$,][]{Straatman14}, hence some of the quiescent galaxies could be star-forming. Cosmic variance is significant ($\sim30\%$). Highly obscured massive star-forming galaxies might also be missed by near-IR surveys \citep[e.g.][]{Daddi09,Caputi12}, although the abundance and redshift distribution of such galaxies is still very uncertain. Finally, extended ($r>3$kpc) galaxies with small $n_{sersic}$ and low surface brightness are more difficult to detect than compact galaxies {(e.g. Trujillo et al. 2006)}.

We caution that the light profiles measured here may not be representative of the stellar mass distribution due to color gradients, with rest-frame UV sizes larger than rest-frame optical sizes. This would imply that the size evolution is stronger. However, using  control sample at $z\sim3$, we find no difference between UV and optical, consistent with \cite{vanderWel14}, who show this effect is $\lesssim10\%$ at $z\sim2$ and decreasing with redshift.

Galaxy sizes may also be overestimated if dust is obscuring a central starburst. Submm sizes of obscured starbursting galaxies could be small: $<1$kpc \citep[e.g.][]{Ikarashi14,Simpson15}. A direct comparison of ALMA submm and rest-frame optical/UV morphologies for the same objects with measured stellar mass will reveal the effect of dust obscuration on UV/optically measured galaxy sizes.

\section{Acknowledgements}
This research was supported by the George P. and Cynthia Woods Mitchell Institute for Fundamental Physics and Astronomy, the National Science Foundation grant AST-1009707 and the NL-NWO Spinoza Grant. Australian access to the Magellan Telescopes was supported through the National Collaborative Research Infrastructure Strategy of the Australian Federal Government. GGK is supported by an Australian Research Council Future Fellowship FT140100933. KEW is supported by an appointment to the NASA Postdoctoral Program at the Goddard Space Flight Center, administered by Oak Ridge Associated Universities through a contract with NASA. We thank Arjen van der Wel, Darren Croton, Duncan Forbes and Alister Graham for useful discussions.


\end{document}